# Green-extraction of carbon thin films from natural mineral Shungite

Anastasia Novikova[1,2,3] and Alina Karabchevsky[1,2,]

[1] School of Electrical and Computer Engineering, Ben-Gurion University of the Negev, Beer-Sheva, 8410501, Israel.

[2] Ilse Katz Institute for Nanoscale Science & Technology, Ben-Gurion University of the Negev, Beer-Sheva, 8410501, Israel.

[3] Engineering School of Non-Destructive Testing of the National Research Tomsk Polytechnic University, Tomsk, 634050, Russia.

## Abstract

Conventional fabrication methods to produce graphene are cumbersome, expensive and are not ecology friendly. This is due to the fact that a large volume of raw materials requires a large number of acids and alkalis, which in turn requires special disposal. Therefore, it is necessary to develop new technologies or refine existing technologies for the production of graphene and create new ecology-safe and effective methods to produce graphene. Here, we utilized the physical dispergation to extract graphene films from natural mineral shungite rock. We studied the structure of shungite via Raman spectrometry and X-ray phases analysis and found that shungite refers to graphite-like mineral structures. From spectral data, we learned that the main constituent of shungite is amorphous carbon in sp2 and sp3 forms. Transmission electron microscopy images of the processed material revealed that the obtained graphene films with well-developed surfaces are as small as 200 nanometers. Our green fabrication method of graphene can be widely used in a variety of fields from electronics (electrodes), optics, biotechnology (biosensors), ecology (sorbents for wastewater, air purification) to list a few.

## Introduction

One of the common and affordable natural materials with catalytic and chemical activity is shungite. Shungite and shungite-bearing rocks are included in a large group of Precambrian carbon-bearing rocks [1]. Shungite rocks are very diverse in the form of manifestations, formation time, genesis and material composition of the ash portion, isotopic composition, aggregate and structural state of shungite carbon. They also differ in physicochemical properties, chemical and mineralogical composition [2]. Shungite is a unique carbon material contained in Precambrian rock of sedimentary origin, the main deposits of which are located in Karelia. Shungite rocks contain carbon in amorphous form (from 5% to 99% depending on the species), minerals (quartz, feldspar, aluminosilicates [3], carbonates, and pyrites), small amounts of bitumen-like organics and water [1]. Five types of shungite rocks are distinguished depending on the carbon content:

Table 1 - Classification of Shungites and their applications [4-12]

| Group | Carbon concentration % | Applications |
| --- | --- | --- |
| I | 99 | graphene production |
| II | 35-75 | Metallurgical industry, in construction |
| III | 25-35 | water purification due to high adsorption properties |
| IV | 10-25 | agriculture, pharmacology, medicine |
| V | 1-10 | agriculture |

Depending on the amount of carbon, types of shungites may differ from each other; at C=64% or less, shungite has a meth-gray variety [13], and at C ≥ 65%, a brilliant variety with a hardness of 3.5-4. Its fracture is a stepped conchoid with adensity of 1.80-2.84 g/cm$^3$ depending on shungite composition; porosity is about 0.5-5%; compressive strength 100-276 MPa; modulus of elasticity (E)-0.31x10$^5$ MPa. Electrical conductivity (1-3)x10$^3$ S/m; thermal conductivity - 3.8 W/m*K. The average value of the coefficient of thermal expansion is in the temperature range from +20 to +600°C is 12x10$^{-6}$K$^{-1}$. Calorific value 7500 kcal/kg. Some types of shungites have reducing, adsorption and catalytic properties, while other types have bactericidal properties.

By its physical and mechanical properties (strength, abrasion, bulk density) shungite is close to the traditionally used filtering materials such asquartz sand. Having a sorption capacity for a wide range of organic substances (surfactants, alcohols, resins, pesticides, petroleum products, etc.), it exhibits specific activity in eliminating particles of a radical nature from water (organochlorines, dioxins). Shungite shows the ability to disinfect and purify water from bacteria, spores, simple microorganisms, blue-green algae. Due to its high catalytic and reducing properties, shungite is often used for wastewater treatment as adsorbents [14,15]. The hybrid structures of shungite consist of sp1+sp2+sp3 forms of hybridized carbon atoms.

Mostly, sp2+sp3 forms are present in a small fraction of sp1. There is little theoretical work on the production of sp1+sp2+sp3, since the probability of obtaining sp3+sp2+sp1 forms of carbon is estimated to be lower as compared to the hybrid phases of other classes [16]. Hybrid structures consist of sp3+sp2+sp1 hybridized atoms including nano-beads [17], glassy carbon [18], nano-flowers, nanotubes functionalized with fragments of graphene layers and carboxylic chains. In addition to the abovementioned structures, there are also minerals in nature in which sp3+sp2+sp1 hybridized structures are present, including chaoites and shungites.

Here, we report on the production of graphene from shungite rock as illustrated in Fig. 1, since shungite contains sp3+sp2+sp1 fungridized graphene-like carbon [19]. There are various ways to obtain graphene plates from shungite, but mostly, they use either heat treatment or treatment with a mixture of acids and alkalis in different concentrations and ratios [20,21]. Here, we propose a new method of shungite processing. We report on the fabrication of graphene from the natural mineral shungite of the Zazhoginsky deposit (Karelia), which belongs to the shungite of group I (the concentration of amorphous carbon is more than 98%) that have adsorption, catalytic and bactericidal properties. Graphene finds applications in electronics, optics, and biotechnology [22,23]. It may find application in other industries [24-33].

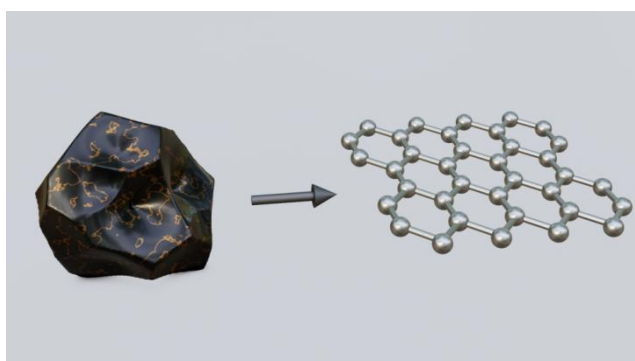

Figure 1 - Illustration of Shungite rock and extracted graphene layer from it.

## Materials and Methods

### Study of the physical and chemical properties of shungite

Samples from the Zazhoginsky field during the geological processes were secondarily deposited from the primary veins and are in the form of pellets. The Zazhogino deposit is also located near the volcanic body, which indicates an additional high-temperature (up to 500°C) heating of shungite rocks .We investigated the surface of the shungite, the presence of mineral inclusions and the composition of the elements. The studies were carried out using a scanning electron microscope (SEM) with a low vacuum and a Quanta 200 SEM tungsten electron source. Its four quadrant displays simultaneously provide surface and phase information via real-time secondary electron (SE) and back-scattered electrons (BSE) images. The Quanta SEM electron microscope was also equipped with an energy-dispersive X-ray spectroscopy (EDS) system for elemental analysis. For the analysis of shungite, we also used the X-ray method. The studies were carried out using a PANalytical Empyrean multipurpose diffractometer for the analysis of powder and solid substances, nanomaterials, thin films, and suspensions. The instrument is equipped with two position-sensitive detectors : X'Celerator 1D and PIXcel-3D for high-speed data acquisition. We study the surface area, radius, and pore volume of shungite using a gas sorption analyzer of the NOVAtouch™ series, analysis of the surface area and pore size by the Brunauer–Emmett–Teller (BET) method (the classical method for determining the volume of voids with nitrogen).

**Extraction of graphene from shungite**

Prior to the production of graphene, shungite samples were crushed to pieces, cleaned of visible contaminants, and washed in distilled water for 5 minutes at room temperature of 25°C. Samples of graphene layers were prepared by the dispergation method using a digital ultrasonic cleaner R Technology under normal conditions (temperature 25°C). One gram of shungite was placed in a 50 ml plastic tube, then 25 ml of distilled water was added, closed and placed in a digital ultrasonic cleaner R Technology for 2 hours. We used the Transmission Electron Microscope (TEM) JEOL JEM 2100F to evaluate and characterize the samples obtained. The JEM-2100F, a Field Emission gun Transmission Electron Microscope, is a state-of-the-art ultra-high-resolution analytical TEM that is capable of providing high spatial resolution atomic imaging and microstructure analysis of material samples.To assess the structural features, we studied the Raman spectra using a LabRam HR Evolution Horiba Raman spectrometer with an excitation range of 325-785 nm with an ultra-low frequency (ULF) module, which allows one to determine the characteristics of the sample at very low frequencies (532 nm), and also measure anti-Stokes spectra. To study the composition of the elements, we used X-ray photoelectron spectroscopy (XPS). The studies were carried out using the ESCALAB 250, a multifunctional instrument that includes surface-sensitive XPS and Auger electron spectroscopy (AES) analysis methods. In this system, XPS high-speed routine analysis is optimized for large areas. The X-ray source, data acquisition system and sample movement are fully computer-controlled, providing the user with an automatic operation to analyze a large number of samples.

## Results and discussion

Prior to obtaining graphene films, shungite samples were checked for the presence of pores and heavy metal impurities in the sample. To understand the physical and chemical properties of shungite, we studied its surfaces, checked the uniformity and presence of mineral inclusions. The indicators of shungite: elemental composition and the combustion temperature, the sp1+sp2+sp3 forms have characteristic absorption spectra was explored via Raman, X-ray and X-ray photoelectron spectroscopy (XPS).

First, we checked the shungite surface using a scanning electron microscope (SEM) to determine the type of shungite. Figure 2 shows scanning electron microscope images of the surface of shungite.

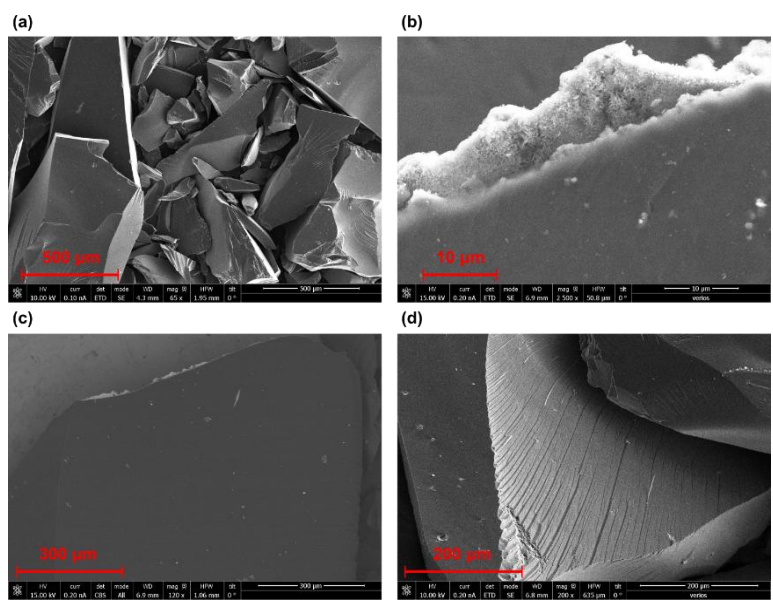

Figure 2 - Scanning electron microscopy images of shungite particles: (a) zoom-offview onparticles at a magnification of x65, (b) shungite surface with characteristic chips particles at a magnification of x200, (c) shungite surfaces with chemical inclusions at a magnification of x120 and (d) shungite surfaces with chemical inclusions at a magnification of x2500.

The shungite is heterogeneous and stepped; on the surface there are characteristic shell spalls of shungite as can be seen from Fig. 2a and Fig. 2b. We found that micropores are present on the surface of shungite as can be seen from (2c and Fig. 2d), but in small amounts. The pores are unevenly distributed over the entire surface and are present only in spongy mineral inclusions (Fig. 2c and Fig. 2d). The mineral contains inclusions of different types and individualized grains and spongy inclusions of micro- and nanometer size, different in shape. Also, mineral inclusions can have a different elemental composition.

Next, we studied shungite samples using the Brunauer-Emmett-Teller (BET) method to determine the number and volume of pores on the surface of shungite to determine the purity of shungite. The presence of a large pore volume may be good for water treatment, but not for the production of graphene films. The obtained pore size values are presented in Table 2.

Table 2 - Shungite pores sizes

| Radius (nm) | Pore volume area ($m^3$/g) | Pore surface area ($m^2$/g) |
|---|---|---|
| 1.49 | 5.40e-06 | 7.26e-03 |
| 1.61 | 1.41e-05 | 1.81e-02 |
| 1.81 | 3.23e-05 | 3.81e-02 |
| 2.04 | 7.17e-05 | 7.67e-02 |
| 2.31 | 9.97e-05 | 1.01e-01 |
| 2.65 | 1.15e-04 | 1.12e-01 |
| 3.09 | 1.34e-04 | 1.25e-01 |
| 3.64 | 1.53e-04 | 1.35e-01 |
| 4.41 | 1.76e-04 | 1.45e-01 |
| 5.60 | 1.98e-04 | 1.54e-01 |
| 7.76 | 2.35e-04 | 1.61e-01 |
| 13.15 | 2.997e-04 | 1.73e-01 |

The total pore radius is 2.04nm, the total pore volume is 0.0003cm$^3$/g and the total pore surface area is 0.173m$^2$/g. Based on the data obtained, it was concluded that shungite has a small number of pores. The pores are unevenly distributed over the surface and make up a small percentage of the total surface of the studied mineral. Pores are present only in mineral inclusions, which is confirmed by the SEM images shown in Fig. 2, but the presence of closed pores is also possible [34].

After detecting inclusions of various shapes on the surface, we investigated their chemical composition of shungite using energy-dispersive X-ray spectroscopy (EDS). We selected random regions in the sungite and selected several areas in each region with following dimensions: $S_1$=300 nm$^2$, $S_2$=270 nm$^2$, $S_{21}$=7200 nm$^2$ and $S_{22}$=900 nm$^2$ (shown in Fig. 3a and 3d). The surfaces and the spectra of the selected areas are shown in Fig. 3.

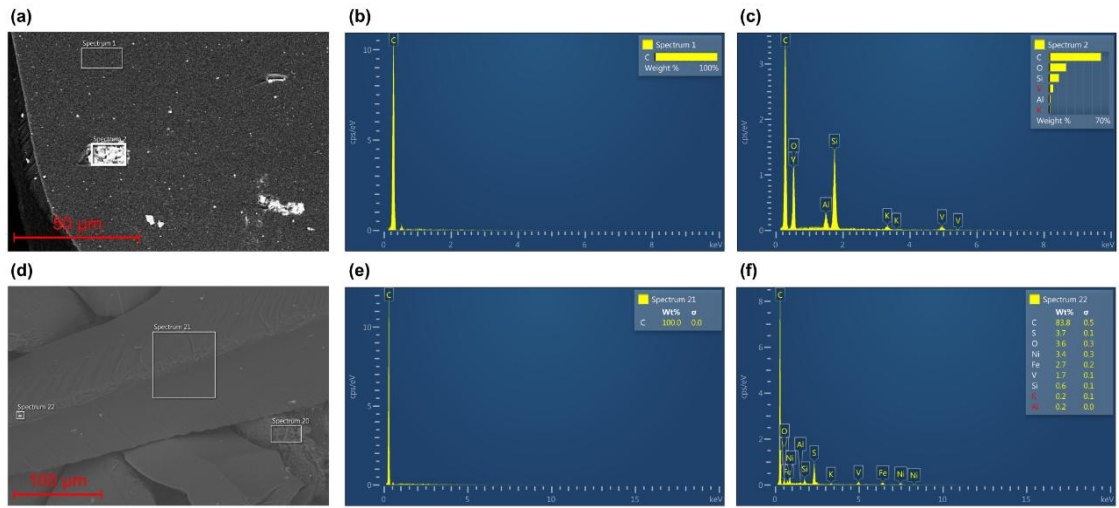

Figure 3 - SEM images and EDS images of shungite particles: (a) studied region 1, (b) chemical composition region 1, (c) chemical composition of region 2, (d) studied region 2, (e) chemical composition region 21 and (f) chemical composition of region 22.

We investigated two regions in each chosen sample as shown in Fig. 3a and Fig.3d. The EDS spectra of region 1 (Fig. 3b) and 21 (Fig. 3e) show that the completely dark sections of the natural mineral are composed of pure carbon (100%). These data confirm that this mineral can be classified as shungite of group I is carbon-rich. The EDS spectra of regions 2 (Fig. 3c) and 22 (Fig. 3f) show that in addition to the small concentrations of carbon, there are contributions of oxygen, silicon, aluminum, nickel, iron and vanadium. It verifies that the sample is shungite.

Next, we examined the elemental composition of the ground sample to identify the main elements of the mineral using the X-ray method. The elements present in shungite inclusions are summarized in Table 3:

Table 3 - The basic elemental composition of the mineral inclusions

| Element | Concentration (wt %) | Element | Concentration (wt %) |
|---------|---------------------|---------|---------------------|
| C       | 55.62               | Al      | 0.86                |
| Si      | 5.55                | Fe      | 12.27               |
| O       | 16.5                | S       | 2.25                |

The main elements of shungite are presented in Table 3 which characterize shungite mineral. In addition, vanadium, potassium, sodium, magnesium, zinc, nickel, molybdenum, and arsenic are also present in small amounts.

The shungite is also characterized by graphite and graphene forms of carbon. Using the X-ray photoelectron spectrometry (XPS) method, we studied the binding of carbon to other elements. The data obtained are shown in Fig. 4.

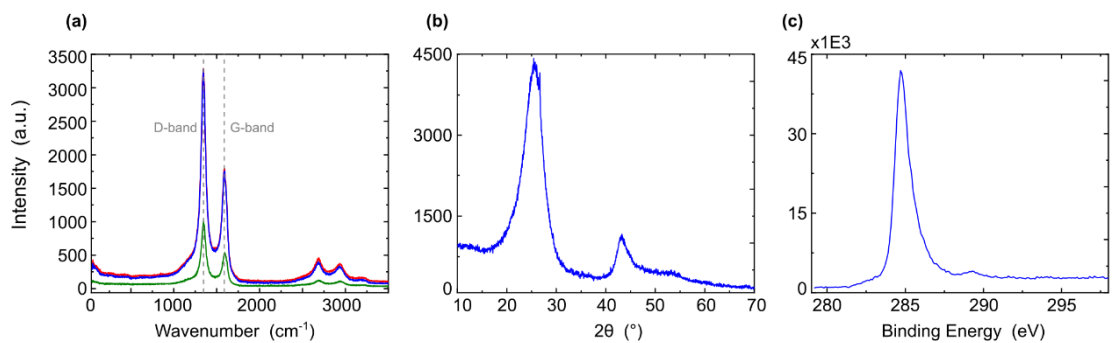

Figure 4 - (a) Raman spectra of natural shungite in different areas, (b) X-ray spectra of shungite and (c) XPS spectra of carbon lines in the shungite spectra.

Figure 4a shows Raman spectra of the studied mineral in different areas. We analysed peaks detected in two spectral regions: 1100-1800 cm$^{-1}$ and 2550-3100 cm$^{-1}$. The peaks that appear in the range of 1100-1800 cm$^{-1}$ are of the first order and can be associated with the peaks of graphite forms of carbon. The peaks that appear in the range of 2550-3100 cm$^{-1}$ are of the second order and are also associated with graphite [35]. The band that appears at 1600 cm$^{-1}$ can be associated with the G-band which appears due to the tangential stretching vibrations of carbon atoms in the hexagons of graphene planes. It appears in the Raman spectra of carbon materials with sp2 bonds, in our case, graphene. The band located at 1330 cm$^{-1}$ is the D-band which appears in the presence of diamond-like sp3 -bonds. In our case, it corresponds to the amorphous structural state of carbon (graphite). The location of those peaks indicates that the main part of amorphous carbon in shungite rock is graphite. The ratio of the intensities of the D (graphite) and G (graphene) bands is traditionally used to assess the degree of ordering of carbon materials. Here, the calculation was performed according to the height of the observed peaks (absolute maxima) of the intensity for D1=1343 cm$^{-1}$, D2=1345 cm$^{-1}$, D3=1345 cm$^{-1}$, for G1=1584 cm$^{-1}$, G2=1585 cm$^{-1}$, G3=1590 cm$^{-1}$. From the ratio of the intensities of the D and G bands (Fig. 4a), it can be seen that thestructure of shungites is more disordered. The second-order peaks also differ in range. The offset as shown in the (Fig. 4a), represents three different peaks 2550-3100 cm$^{-1}$, D1"= 2657 cm$^{-1}$, D2"=2657 cm$^{-1}$, D3"=2654

cm$^{-1}$, G1"=2913 cm$^{-1}$, G2"=2938 cm$^{-1}$, G3"=2930 cm$^{-1}$, which also indicates the disorder of the structure of shungite. Figure 4b shows X-ray spectra shungite, wide spectra at 26 degrees refer to carbon in its amorphous form, and more specifically, to graphite-like carbon, as well as a secondary peak at 42 degrees. Based on the intensity of X-ray peaks we confirm that the main form of shungite is indeed the amorphous carbon and that the studied shungite sample belongs to group I.

To determine the numbers of sp2 and sp3 carbon bonds, we studied the samples using X-ray photoelectron spectrometry. Data are presented in Table 4 and in Fig. 4c.

Table 4 - Elements Identifier and Quantitative assessment of shungite before dispergation

| Name | Peak (BE) | FWHM (eV) | Area (CPS·eV) | Atomic % |
| --- | --- | --- | --- | --- |
| C1S scan A *sp3* C-C | 284.32 | 0.59 | 7059.83 | 13.25 |
| C1S scan D *sp2* C=C | 284.75 | 0.66 | 26502.12 | 49.73 |
| C1S scan B C=O | 285.31 | 0.69 | 9135.39 | 17.14 |
| C1S scan C=C-O | 285.85 | 1.39 | 10594.59 | 19.88 |

Figure 4c shows the XPS spectra in which the intense peak corresponds to an energy of 284.75 eV and can be attributed to carbon. The full width at half maximum (FWHM) of the spectral line is 0.66 eV. Table 4 shows that in addition to the C-C and C=C bonds of carbon, shungite also contains C=O and C=C-O bonds. This data confirms results from X-ray and Raman spectroscopies shown in Fig. 4b.

After examining the shungite samples, we processed the samples via mechanical method - dispergation. Since large volumes of samples can be used with this method, there is no secondary contamination and there is no need to use subsequent sample processing. Figures 5a-b show the surface areas of the dispergated shungite examined under the transmission electron microscope (TEM). The surface of the samples has changed as compared to Fig. 2. To determine the numbers of sp2 and sp3 carbon bonds, we studied the samples using Raman and XPS spectroscopies as shown in Fig. 5 and Table 4.

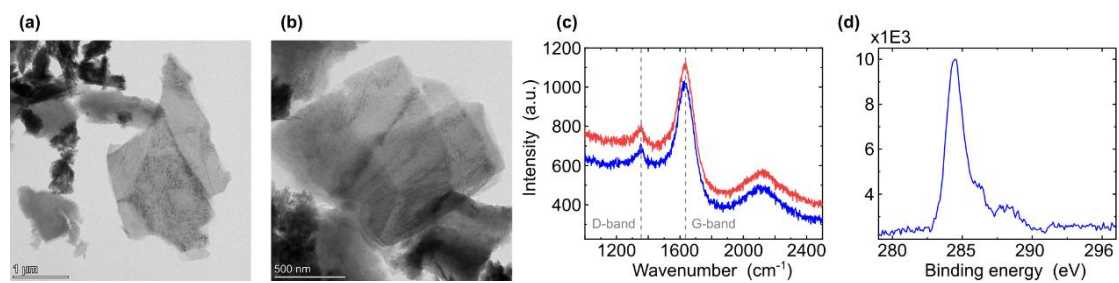

Figure 5 - TEM images of dispergated shungite sample at (a) magnification of X13500 and (b) magnification of X35000. (c) Raman spectrum of processed shungite and (d) XPS spectra of lines of carbon in the shungite spectra.

Figures 5a-b show TEM images of dispersed shungite sample. It shows that the dispergation process crushes the particles to form thin films with high specific surfaces. Since our shungite consists mainly of carbon in an amorphous form, we conclude that films observed in Fig. 5a-b are of graphene-like nature.

Figure 5c shows Raman spectrum of dispersed shungite sample. An analysis was made of the peaks recorded in two spectral regions: 1100-1800 cm$^{-1}$ and 2550-3100 cm$^{-1}$. From the obtained Raman spectra, we can conclude that the spectra of the treated shungite are different. They do not have second-order peaks and the intensity of the D and G lines has changed, peak position is D1=1359.82 cm$^{-1}$, D2=1359.82 cm$^{-1}$, G1=1639.89 cm$^{-1}$, G2=1619.98 cm$^{-1}$. The ratio of the peaks has also changed indicating the structural changes that have occurred in the samples. In untreated shungite, the D peak related to sp3 bonds (graphite) prevailed; in the treated sample, it greatly decreased. The G line related to sp2 bonds (graphene) has expanded. The peaks have moved to the right.

Table 5 - Elements Identifier and Quantification after the dispergation

| Name | Peak (BE) | FWHM (eV) | Area (CPS·eV) | Atomic % |
| --- | --- | --- | --- | --- |
| C1S scan C *sp3* C-C | 284.21 | 0.84 | 1645.01 | 16.29 |
| C1S scan A *sp2* C=C | 284.79 | 1.44 | 8709.01 | 53.63 |
| C1S scan D C-O | 286.22 | 1.55 | 3098.99 | 19.09 |
| C1S scan E O-C=O | 288.46 | 1.66 | 1783.84 | 10.99 |

Figure 5d and Table 5 show XPS spectra and data of dispergatedshungite sample. The graph shows the peak of the sp2 form of carbon, located at 284.79 eV, which is attributed to carbon.

The full width at half maximum of the spectral line (FWHM) was 1.44 eV. Comparing the percentage of sp2 and sp3 forms, we can conclude that there are more sp2 forms. The percentage of sp3 forms has increased and the percentage of C=C-O has decreased by 8.89% while the C=O bonds have disappeared. In addition, C-O bonds have appeared.

## Conclusions

We considered the production of graphene films from the natural mineral shungite using a mechanical method - dispergation. This is a promising method for producing industrial volumes of graphene and graphene-like matter. Reported by us method is efficient and easy to use. It does not require additional costs and does not imply secondary contamination of the environment and this creates new prospects for the use of Shungite.

Next, we conducted research on processed shungite. TEM images of the processed material showed that the resulting carbon films with a well-developed surface were only 200 nanometers in size. Raman spectra of processed shungite showed that films are composed of carbon in sp2 form – graphene. XPS data showed that the percentage of sp2 forms of carbon is increased. We obtained undeformed graphene plates by a method that no one had ever used before. The outcomes of this research open the door to the new and environmentally safe method of graphene production. Green fabrication of graphene can be used in a variety of fields from electronics (electrodes), optics to biotechnology (biosensors, drug delivery, bio-medicine), energy (supercapasitors, solar cells, energy harvesting) and ecology (sorbents for wastewater, air purification).

## Declaration of competing interest

The authors declare that they have no known competing financial interests or personal relationships that could have appeared to influence the work reported in this paper.

## Acknowledgments

We are thankful to Israel Science Foundation (ISF) grant No. 2598/20 for supporting our research.We thank Aviad Katiyi for the fruitful discussions and help in the preparation of the manuscript.